\documentclass[letterpaper,12pt]{article}

\usepackage{fullpage}
\usepackage{latexsym}
\usepackage{graphicx}

\usepackage{amsthm}
\usepackage{amsmath}
\usepackage{amsfonts}
\usepackage{amssymb}

\usepackage{bm}

\usepackage{algorithm}
\usepackage{algpseudocode}

\graphicspath{{./figures/}}

\newtheorem{theorem}{Theorem}
\newtheorem{lemma}[theorem]{Lemma}
\newtheorem*{conjecture}{Conjecture}

\usepackage{url}

\newcommand{\functionname}[1]{\ensuremath{\text{\sf #1}}}

\newcommand{\dest}{\functionname{dest}}
\newcommand{\vertex}{\functionname{vert}}
\newcommand{\move}{\functionname{km}}
\newcommand{\cell}{\functionname{cell}}
\newcommand{\freq}{\functionname{f}}
\newcommand{\rfreq}{\functionname{r}}
\newcommand{\relative}{\functionname{rel}}

\newcommand{\knightsmoves}{\ensuremath{\mathfrak{K}}}

\usepackage[dvipsnames]{xcolor}
\usepackage{hyperref}
\hypersetup{
  colorlinks   = true, 
  urlcolor     = Blue,
  linkcolor    = ForestGreen, 
  citecolor    = Bittersweet 
}

\newcommand{\keywords}[1]{
  \small\textbf{Keywords:} #1
}

\title{Tourneys and the Fast Generation and Obfuscation of Closed Knight's Tours
(Preliminary Version)}

\author{
  Ian Parberry\\
  Dept.\ of Computer Science \& Engineering\\
  University of North Texas\\
	Denton, TX, USA\\
	\url{http://ianparberry.com}}

\begin {document}
\maketitle

\begin{abstract} 
New algorithms for generating closed knight's tours 
are obtained by generating a vertex-disjoint cycle cover of the knight's graph and joining the resulting cycles. It is shown experimentally that these algorithms are significantly faster in practice than previous methods. A fast obfuscation algorithm for closed knight's tours that obscures obvious artifacts created by their method of generation is also given, along
with visual and statistical evidence of its efficacy. 
\end{abstract}



\section{Introduction}
\label{sec:intro}

A {\em closed knight's tour\/} is a sequence of moves for a single knight that returns the knight to its start position after visiting every square of a finite rectangular chessboard exactly once.
It is said that Euler~\cite{Euler1759} was in 1759 the first person to attempt the construction of a closed knight's tour on the standard $8 \times 8$ chessboard using a random walk algorithm. Since then the problem has attracted a great deal of interest.
We will, for convenience, abbreviate {\em closed knight's tour\/} to {\em knight's tour}.

There are three primary methods for constructing knight's tours; random walk, neural network, and divide-and-conquer.  The random walk and neural network algorithms create a different knight's tour every time they are run, but require exponential time. The divide-and-conquer algorithm creates the same knight's tour every time it is run, but its running time is linear in the size of the board. Aesthetically, the knight's tours created by random walk and neural network are pleasing to the eye because they are unstructured and chaotic, and those created by divide-and-conquer are pleasing to the eye for the completely opposite reason, because they have a structured and regular appearance.

Define a {\em tourney}\footnote{So named because a medieval tourney can be viewed as a collection of knights riding in closed loops.} of size $k \geq 1$ to be a collection of non-trivial sequences of moves for $k$ knights that returns each knight to its start position after every square of a finite rectangular chessboard has been visited by exactly one knight exactly once. The {\em size\/} of a tourney is the number of knights. A closed knight's tour is a tourney of unit size. Tourney generation and its applications has until now gone largely unstudied by the academic community.

We describe some fast algorithms for constructing large, structured tourneys deterministically, and
give experimental evidence that the two standard methods for generating random knight's tours (random walk and neural networks) can be modified to generate tourneys instead with a significant decrease in run time. When combined with a fast algorithm for creating knight's tours from tourneys, this gives us a faster method of generating random knight's tours.
The three knight's tours generation algorithms described above generate knight's tours with visual artifacts that betray their method of construction. We describe a fast obfuscation algorithm that obscures these artifacts.

The CPU times reported in this paper were for a
C++ implementation compiled using Microsoft\textsuperscript{\textregistered} 
Visual Studio 2019\textsuperscript{\textregistered} and executed on an Intel\textsuperscript{\textregistered} Core\textsuperscript{\texttrademark} i9-7980XE
under Microsoft\textsuperscript{\textregistered} Windows 10\textsuperscript{\textregistered}.
Links to the cross-platform open-source code and accompanying documentation can be found in the Supplementary Material (see Section~\ref{sec:supplement}).

The remainder of this paper is divided into six sections,
Section~\ref{sec:definitions} covers notation and definitions.
Section~\ref{sec:prior} covers prior work on the generation of knight's tours.
Section~\ref{sec:rails} introduces the concepts of {\em rail\/} and {\em rail switching}.
Section~\ref{sec:join} describes the {\em join\/} algorithm for tourneys, which switches a set of edge-disjoint rails in a spanning tree of a multigraph called the {\em rail graph}.
Section~\ref{sec:tourneys} contains some new tourney generation algorithms.
Section~\ref{sec:shatter} covers the {\em shatter\/} algorithm for tourneys based on switching a pseudo-random set of edge-disjoint rails, and its application to obfuscating knight's tours.
After a brief conclusion in Section~\ref{sec:conclusion},
Section~\ref{sec:supplement} contains URLs for larger diagrams, the full data set, and open-source code that can be used to verify the claims made in this paper.


\section{Notation and Definitions}
\label{sec:definitions}

\begin{figure}[hbt]
  \centering
    \includegraphics[height=1.25in]{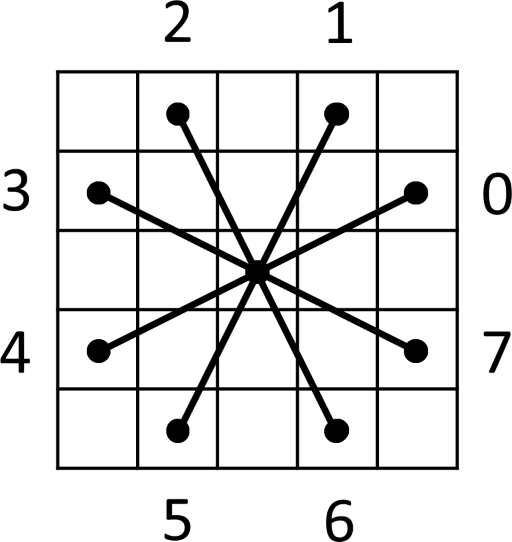}
		\hspace{0.2in}
    \includegraphics[height=1.25in]{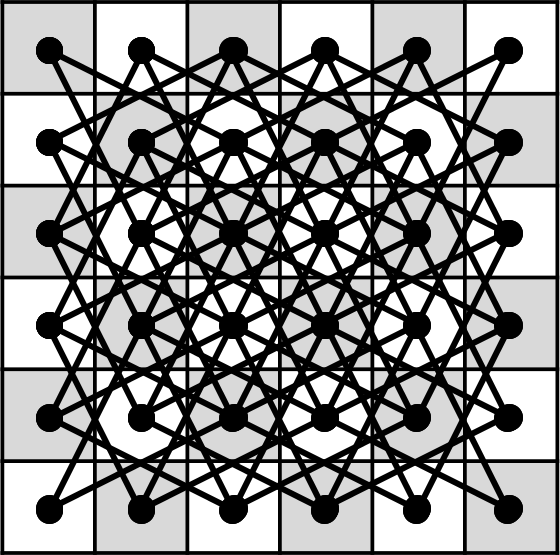}
  \caption{Left: The knight's moves from the center cell numbered counterclockwise 0 through 7.
		Right: The knight's graph $K_6$ on a $6 \times 6$ chessboard.}
  \label{fig:knightsgraph}
\end{figure}

A knight in the game of chess has 8 possible moves available to it, numbered 0 through 7 in 
Figure~\ref{fig:knightsgraph} (left).
For convenience, define $\knightsmoves = \{0,1,\ldots,7\}$.
For all $n \in \mathbb{N}$,
the $n \times n$ {\em knight's graph\/} $K_n$ is a labeled undirected bipartite graph with $n^2$ vertices, one for each square (also called a {\em cell\/}) of an $n \times n$ chessboard, and an edge between vertices $u$ and $v$ iff a knight can move from cell $u$ to cell $v$.
$K_1$, and $K_2$ have zero edges. 
$K_3$ has 8 edges and degree 2. $K_4$ has 24 edges and degree 4. For $n \geq 5$, $K_n$ has $4n^2 - O(n)$ edges and degree 8. $K_n$ is completely connected iff $n \geq 4$.
Figure~\ref{fig:knightsgraph} (right) shows $K_6$.

Suppose $K = (V, E)$ is a knight's graph. 
Define the function $\dest: V \times \knightsmoves \mapsto  V$ such that
for all $v \in V$, $i \in \knightsmoves$, $\dest(v, i)$ is the vertex reached
by move $i$ from $v$ if it exists, and is undefined otherwise.
Conversely, define $\move: V \times V \mapsto \knightsmoves$ as follows. 
For all $u, v \in V$, $\move(u,v)$ is the move that takes a knight from $u$ to $v$
if one exists, and is undefined otherwise.
Suppose $V = \{v_0, v_0, \ldots, v_{n^2 - 1}\}$.
Number the cells of an $n \times n$ chessboard in row-major order
from 0 to $n^2 - 1$.  
Define the functions $\vertex : \{0,1,\ldots,n^2 - 1\} \mapsto V$
such that for all $0 \leq i < n^2$, $\vertex(i) = v_i$, and
$\cell : V \mapsto \{0,1,\ldots,n^2 - 1\}$
such that for all $v \in V$, $\cell(v_i) = i$.

Adjacency in a knight's graph $K = (V, E)$ can be tested using a small number of 
arithmetic operations, since $\vertex(i)$ is in row 
$y_i = \lfloor i/n \rfloor$ and column $x_i = i \bmod n$. 
Let $\mathcal{M} = \{(\pm x, \pm y) \mid x,y \in \{1,2\} \mbox{ and } x + y = 3\}$
be the set of horizontal and vertical displacements for the eight possible knight's moves. Then, for all $0 \leq i,j < n$, $\vertex(i)$ is adjacent to $\vertex(j)$ iff
\begin{displaymath}
  j \in \mathbb{Z}_n \cap 
	\{(y_i + y) n + x_i + x \mid
     0 <  x_i + x < n \mbox{~and~}
     (x, y) \in \mathcal{M}
  \}.
\end{displaymath}

A {\em knight's tour\/} is a Hamiltonian cycle on a knight's graph. It is well-known that knight's tours exist on $n \times n$ chessboards for all even $n \geq 6$.
A {\em tourney\/} is a vertex-disjoint cycle cover of the knight's graph, that is, a set of cycles on the knight's graph such that every vertex of the graph is in exactly one cycle. The {\em size\/} of a tourney is the number of cycles that it contains. A knight's tour is therefore a tourney of unit size.
Let $\mathfrak{C}_n$ denote the set of closed knight's tours and $\mathfrak{T}_n$ denote the set of tourneys on $K_n$. Clearly, $\mathfrak{C}_n \subset \mathfrak{T}_n$.


\section{Prior Generation Algorithms}
\label{sec:prior}

There are many methods for generating closed knight's tours dating back to Euler's algorithm~\cite{Euler1759}, which consists of a knight taking a random walk on the chessboard until it either ends up back at the start cell with all cells having been visited, or something close to it that can be patched up by hand by a observer possessed of sufficient perspicacity (see Ball and Coxeter~\cite{Ball74Mathematical} for more details). While it was possible (although tedious) for Euler to run
his algorithm by hand on the regulation $8 \times 8$ board in 1759, Euler's algorithm does not scale well with board size even when the power of current computers has been harnessed. Fortunately, much progress has been made since then. as we will see in the rest of this section.

Computer-generated knight's tours often have visual idiosyncrasies that make it easy to identify the generation algorithm used. Given sufficiently many examples, a statistical analysis of the moves is even more likely to distinguish between generation algorithms.
More formally, for all $i \in \knightsmoves$, 
$\freq_i(T): \mathfrak{C}_n \mapsto [0,1]$ is defined as follows.
For a given $n \times n$ knight's tour $T = (V,E) \in \mathfrak{C}_n$, 
$\freq_i(T)$ is the frequency of move $i$ in $T$, that is,
$$\freq_i(T) = \frac{\lVert \{(u,v) \in E \mid \move(u,v) = i \text{ or } \move(v,u) = i\} \rVert}{2 n^2}.$$
The {\em move distribution\/} of $T$ is defined to be the sequence 
$\freq_0(T), \freq_1(T),\ldots,\freq_7(T)$.
Note that
$$\sum_{i \in \knightsmoves}\freq_i(T) = 1.$$
Similarly, if $\mathfrak{T} \subseteq \mathfrak{C}_n$
define $\freq_i(\mathfrak{T})$ to be the number of occurrences of move $i \in \knightsmoves$ in $\mathfrak{T}$, that is,
$$\freq_i(\mathfrak{T}) = \sum_{T \in \mathfrak{T}} \frac{\freq_i(T)}{\lVert \mathfrak{T} \rVert}.$$
The {\em move distribution\/} of $\mathfrak{T}$ is then defined to be the sequence 
$\freq_0(\mathfrak{T}), \freq_1(\mathfrak{T}),\ldots,\freq_7(\mathfrak{T})$.
Note that
$$\sum_{i \in \knightsmoves}\freq_i(\mathfrak{T}) = 1.$$

Knight's tours constructed using three of the most practical
knight's tour generation algorithms, Warnsdorff's algorithm,
neural networks, and divide-and conquer,
will be compared and contrasted visually and statistically in the following three subsections.

\subsection{Warnsdorff's Algorithm}

\begin{figure}[htb]
  \centering
    \includegraphics[scale=0.6]{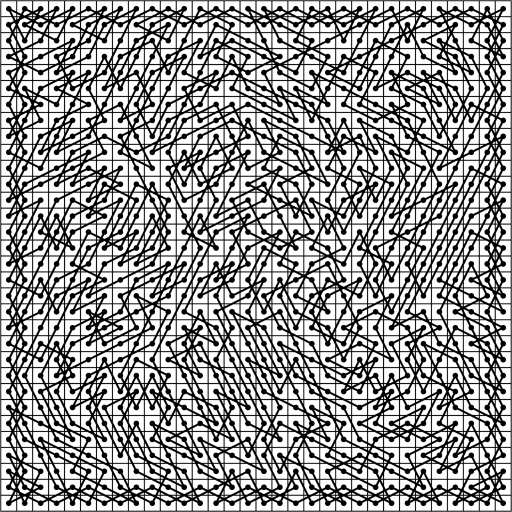}   
		\caption{A $32\times 32$ knight's tour generated by Warnsdorff's algorithm.}
  \label{fig:warnsdorff32}
\end{figure}

\begin{figure}[htb]
  \centering
    \includegraphics[scale=0.44]{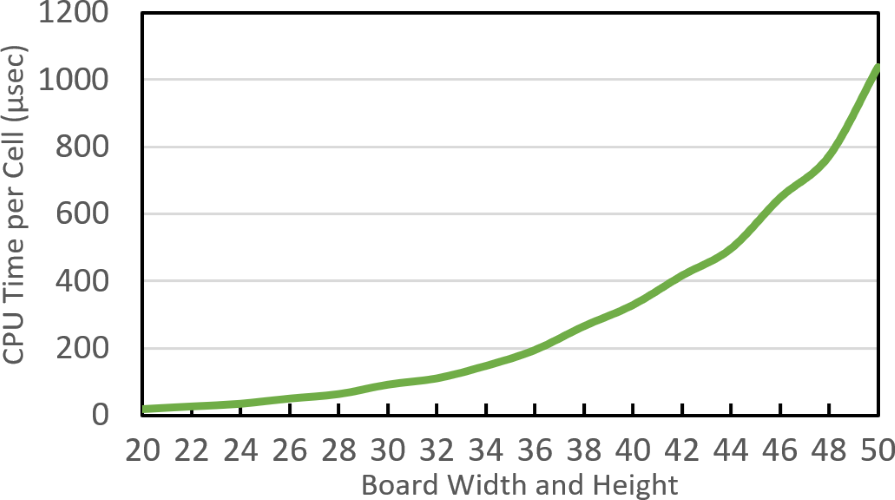}   
		\caption{CPU time per cell averaged over 1,000 $n \times n$ knight's
		tours generated by Warnsdorff's algorithm for even $20 \leq n \leq 50$.}
  \label{fig:time-wt}
\end{figure}

Warnsdorff
(see Conrad {\em et al.\/}~\cite{Conrad1992, Conrad1994}) introduced an heuristic that renders Euler's random walk more practical: Instead of making a random knight's move, make a move randomly chosen from the set of moves that have the minimum number of moves leaving them. Although this is counter-intuitive, it appears to work in practice.
Warnsdorff's heuristic creates tours that have a marked tendency to run in parallel lines, as can be seen in the $32 \times 32$ example shown in Figure~\ref{fig:warnsdorff32}. The running time of Warnsdorff's version appears to increase exponentially with board size (see Figure~\ref{fig:time-wt}). 

\begin{figure}[htb]
  \centering
	  \hspace*{\fill}
    \includegraphics[scale=0.42]{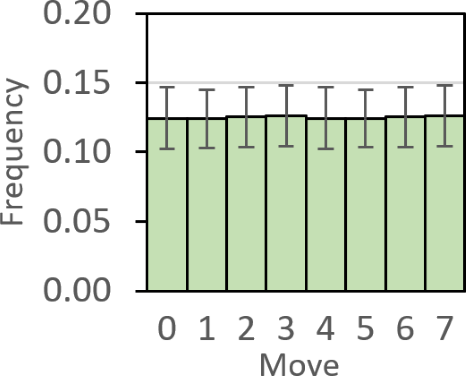} 
	  \hspace*{0.3in} 
    \includegraphics[scale=0.42]{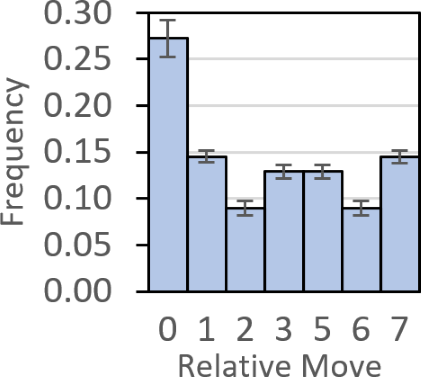} 
	  \hspace*{\fill}
		
		\caption{Move distribution (left) and relative move distribution (right) for 1,000 pseudo-random
		$50 \times 50$ knight's tours generated by Warnsdorff's algorithm. The error bars show
		$\pm 1$ standard deviation.}
  \label{fig:movedist-wt}
\end{figure}	

Figure~\ref{fig:movedist-wt} (left) shows the move distribution for 1,000 pseudo-random
$50 \times 50$ knight's tours generated by Warnsdorff's algorithm, which is close to the uniform distribution. Notice, however that the standard deviation (shown by the error bars) is very large
because Warnsdorff's heuristic tends to amplify any small discrepancy between move frequencies. 
This means that there can be a large amount of variation between the move distributions of individual knights tours. However, there is a much more striking method of identifying knight's tours generated by Warnsdorff's algorithm. A close examination of Figure~\ref{fig:warnsdorff32} reveals that knight's moves are repeated (that is, the relative move is 0) more often that one might expect in a random knight's tour.

More formally, given a pair of moves $i,j \in \knightsmoves$, we say that move $j$ {\em relative to\/} move $i$ is $\relative(i,j) = i - j \pmod{8} \in \knightsmoves$. Therefore, for example, if $\relative(i,j) = 0$, then $i = j$, and if $\relative(i,j) = 4$, then $i$ is the exact opposite move to $j$.
For all $i \in \knightsmoves$, the {\em relative frequency function\/}
$\rfreq_i: \mathfrak{C}_n \mapsto [0,1]$ is defined as follows.
For a given $n \times n$ knight's tour $T = (V,E) \in \mathfrak{C}_n$, 
define $\rfreq_{i}(T)$ to be
\begin{displaymath}
\rfreq_i(T) = \frac
{\lVert \{(u,v),(v,w) \in E \mid u < v,w  \text{ and }
 \relative(\move(u,v) , \move(v,w)) = i\} \rVert}
{2n^2}.
\end{displaymath}
The {\em relative move distribution\/} of $T$ is then defined to be the sequence 
$$\rfreq_0(T), \rfreq_1(T), \rfreq_2(T),\rfreq_3(T),\rfreq_5(T),\rfreq_6(T),\rfreq_7(T).$$
Note that $\rfreq_4(T)$ is not included since it is always equal to zero (in a closed knight's tour no move can be followed by its exact opposite move), and
$$\sum_{i \in \knightsmoves}\rfreq_i(T) = 1.$$
If $\mathfrak{T} \subseteq \mathfrak{C}_n$
define $\rfreq_i(\mathfrak{T})$ to be 
$$\rfreq_i(\mathfrak{T}) = \sum_{T \in \mathfrak{T}} \frac{\rfreq_i(T)}{\lVert \mathfrak{T} \rVert}.$$
The {\em relative move distribution\/} of $\mathfrak{T} \subseteq \mathfrak{C}_n$
is then defined in the obvious manner. Note that
$$\sum_{i \in \knightsmoves}\rfreq_i(\mathfrak{T}) = 1.$$

Figure~\ref{fig:movedist-wt} (right) shows the relative move distribution for 1,000 pseudo-random
$50 \times 50$ knight's tours generated by Warnsdorff's algorithm. As expected, the most frequent relative move is move 0 at 27.2\,\%, which represents a repeat of the previous move.
The next most frequent relative moves are moves 1 and 7 (back and to either side of the
previous move) at 14.5\,\%,
followed by 3 and 5 (forward and to either side of the
previous move) at 12.9\,\%, and 2 and 6 (orthogonal to the
previous move) at 9\,\%.
which represents a repeat of the previous move.
This is consistent with Warnsdorff's heuristic creating sequences of repeated moves
(relative move 0), preferentially staying close to them (relative moves 1, 3, 5, 7) rather than
branching orthogonally to them (relative moves 2, 6).


\subsection{Neural Networks}

\begin{figure}[htb]
  \centering
    \includegraphics[scale=0.6]{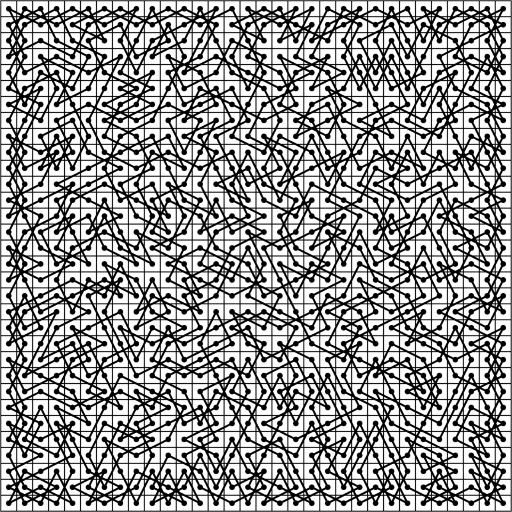}   
		\caption{A $32\times 32$ knight's tour generated by the Takefuji-Lee neural network.}
  \label{fig:neural32}
\end{figure}

The neural network of Takefuji and Lee~\cite{Takefuji92Neural} appears to almost always require exponential time to converge, and Parberry~\cite{ParberryKnight1996} provided experimental evidence that it is significantly slower than Warnsdorff's algorithm.
Figure~\ref{fig:neural32} shows a $32\times 32$ knight's tour generated by the Takefuji-Lee neural network. A visual comparison of Figure~\ref{fig:neural32} with Figure~\ref{fig:warnsdorff32} appears to show that the neural network does not share the Warnsdorff's heuristic's tendency to repeat moves.

\begin{figure}[htb]
  \centering
	  \hspace*{\fill}
    \includegraphics[scale=0.42]{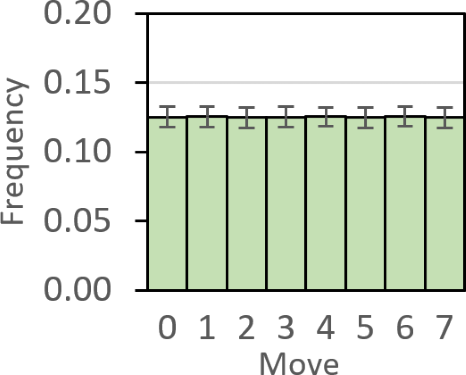} 
	  \hspace*{0.3in} 
    \includegraphics[scale=0.42]{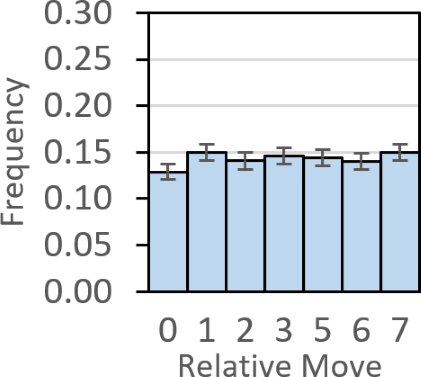} 
	  \hspace*{\fill}
		
		\caption{Move frequency (left) and relative move frequency (right) for 1,000 pseudo-random
		$50 \times 50$ knight's tours generated by the Takefuji-Lee neural network. The error bars show
		$\pm 1$ standard deviation.}
  \label{fig:movedist-tlt}
\end{figure}	

Figure~\ref{fig:movedist-tlt} shows the move distribution and relative move distribution for 1,000 pseudorandom $40 \times 40$ knight's tours generated by the Takefuji-Lee neural network.
The move distribution is, if anything, even more uniform than the move distribution of Warnsdorff's algorithm in Figure~\ref{fig:movedist-wt}, but the distinction is not strong enough to distinguish between them.
The relative move distribution, however, does not show the preference for relative move 0 that is shown by Warnsdorff's algorithm, and is therefore a fairly reliable method of distinguishing between the two generation algorithms.


\subsection{Divide-and-Conquer}

\begin{figure}[htb]
  \centering
    \includegraphics[scale=0.6]{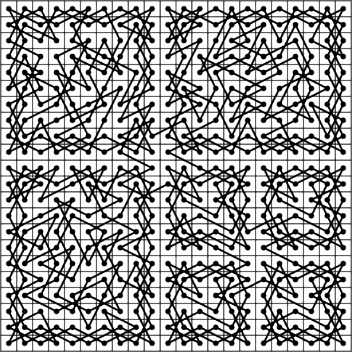}   
		\caption{A $22\times 22$ knight's tour generated by the divide-and-conquer algorithm.}
  \label{fig:dc22}
\end{figure}

The divide-and-conquer algorithm of Parberry~\cite{ParberryKnight1997} is a deterministic algorithm (that is, it generates the same knight's tour every time it is run) that uses $O(n^2)$ time (which is linear in the number of cells).
It creates highly-structured tours that can easily be distinguished by eye, for example, a $22 \times 22$ knight's tour is shown in Figure~\ref{fig:dc22}.


\section{Rails}
\label{sec:rails}

A {\em rail\/} in a subgraph $G = (V, E)$ of a knight's graph $K$ consists of a pair of parallel moves between cells that are separated by knights moves that are not present in $G$, that is, an unordered pair of edges $r=(e, e')$ such that $e = (v_0, v_1)\in G$, $e' = (v_2, v_3)\in G$, and
$(v_0, v_2), (v_1, v_3) \in K \setminus G$.
Suppose $v_0 < v_1$. $v_2 < v_3$, $v_0 < v_2$, and $v_1 < v_3$.
We will call $\move(v_0, v_1) = \move(v_2, v_3) \in \{4,5,6,7\}$ the {\em primary move\/} of $r$, 
and $\move(v_0, v_2) = \move(v_1, v_3) \in \{4,5,6,7\}$ the {\em cross move\/} of $r$.
Note that a rail is completely specified by its topmost vertex, its primary move, and its cross move.

\begin{figure}[htb]
  \centering
    \includegraphics[scale=0.5]{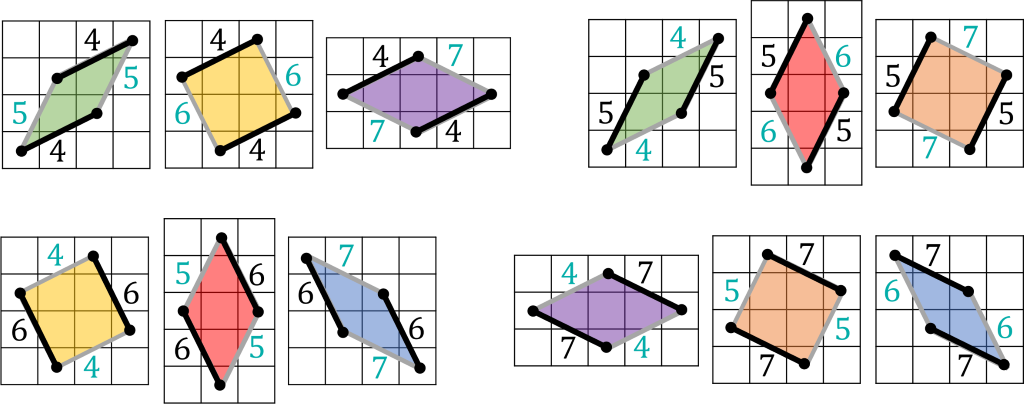}
  \caption{The three rails with primary moves (in row-major order) 4, 5, 6, and 7. The black numbers are primary moves, the cyan numbers are cross moves}
  \label{fig:rails}
\end{figure}

\begin{lemma}
\label{lemma:railcount}
Every move in a subgraph of the knight's graph can be part of at most 6 rails.
\end{lemma}

\begin{proof}
Each downward move (moves 4, 5, 6, and 7) appears as the primary move in 3 types of rail
(see Figure~\ref{fig:rails}), giving 6 distinct occurrences of each move.
\end{proof}

\begin{figure}[ht]
  \centering
    \includegraphics[scale=0.5]{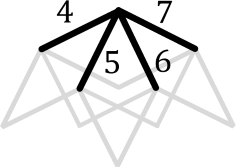}
  \caption{The potential downwards moves from any cell are moves 4--7.
	The outlines of the potential downward rails that use these as the primary move are shown
	overlapping in gray.}
  \label{fig:bird4-7}
\end{figure}

\begin{figure}[ht]
  \centering
    \includegraphics[scale=0.5]{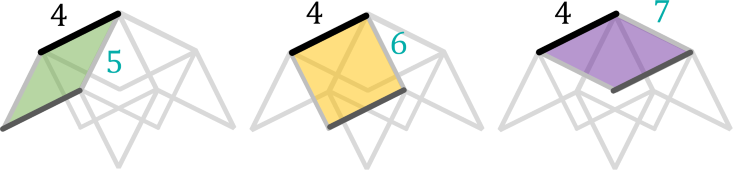}
  \caption{A rail with primary move 4 has cross moves $5, 6, 7$.}
  \label{fig:bird4}
\end{figure}

\begin{figure}[ht]
  \centering
    \includegraphics[scale=0.5]{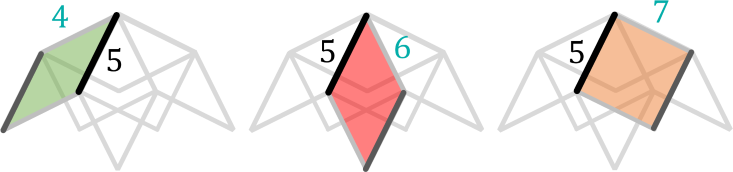}
  \caption{A rail with primary move 5 has cross moves $4, 6, 7$.}
  \label{fig:bird5}
\end{figure}

\begin{figure}[ht]
  \centering
    \includegraphics[scale=0.5]{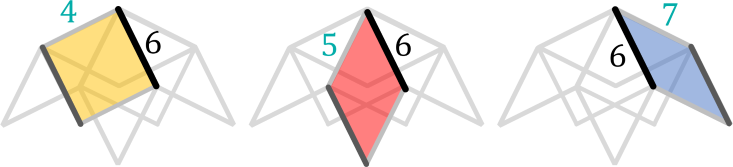}
  \caption{A rail with primary move 6 has cross moves $4, 5, 7$.}
  \label{fig:bird6}
\end{figure}

\begin{figure}[ht]
  \centering
    \includegraphics[scale=0.5]{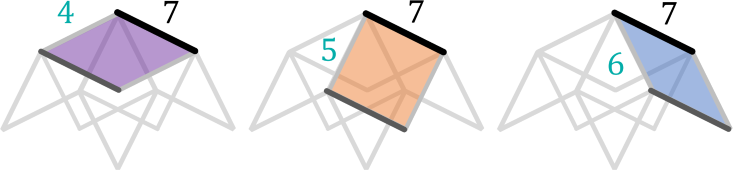}
  \caption{A rail with moves 7 has cross moves $4, 5, 6$.}
  \label{fig:bird7}
\end{figure}

\begin{algorithm}[htb]
\begin{algorithmic}[1]
\Function{FindRails}{$G$} \Comment $G=(V,E)$ is a subgraph of $K_n$ for some $n \geq 4$
\State $S \gets \{\}$ \Comment $S \subseteq E \times E$ is the set of rails found so far
\For{$u \in V$} \Comment for each vertex $u$ of $G$
  \For{$e = (u,v) \in E$ such that $\move(u, v) \geq 4$} \Comment for each downward edge from $u$
	     \For{$4 \leq j \leq 7$, $j \not= \move(u,v)$} \Comment $j$ is the cross move
				\State $u' \gets \dest(u, j)$ \Comment $u'$ is cross move $j$ away from $u$
				\State $v' \gets \dest(v, j)$ \Comment $v'$ is cross move $j$ away from $v$
				\State $e' \gets (u',v')$ \Comment $e'$ is the primary move opposite $e$
				 \If{$e' \in E$ and $(u,u'), (v,v') \not\in E$} \Comment the rail $(e, e')$ is present in $G$
				   \State $S \gets S \cup \{(e, e')\}$ \Comment add $(e, e')$ to the rail set
				 \EndIf
       \EndFor
  \EndFor
\EndFor
\State\Return $S$ \Comment $R = \{ (e, e') \mid  (e, e') \mbox{ is a rail in } G\}$
\EndFunction
\end{algorithmic}
\caption{Find all rails in a subgraph of the knight's graph.}
\label{algorithm.find}
\end{algorithm}

\begin{theorem}
\label{theorem:railfind}
The set of rails in a subgraph of $K_n$ can be found in $O(n^2)$ time.
\end{theorem}

\begin{proof}
Suppose $G$ is a subgraph of $K_n$ for some $n \geq 4$.
Consider function {\sc FindRails}$(G)$ described in Algorithm~\ref{algorithm.find}.
The for-loop on Lines~3--14 iterates through the vertices $u \in G$.
The for-loop on Lines~4--13 iterates through the edges $e=(u,v)$
such that $\move(u,v) > 4$, that is, $\cell(v)$ can be reached by a downward move $i$ from 
$\cell(u)$, as shown in Figure~\ref{fig:bird4-7}).
Noting that
a rail with primary move 4 can have cross move 5, 6, or 7 (Figure~\ref{fig:bird4}),
a rail with primary move 5 can have cross move 4, 6, or 7 (Figure~\ref{fig:bird5}),
a rail with primary move 6 can have cross move 4, 5, or 7 (Figure~\ref{fig:bird6}), and
a rail with primary move 7 can have cross move 4, 5, or 6 (Figure~\ref{fig:bird7}),
the for-loop on Lines~5--12 iterates through all of the cross moves $j$ that
can potentially be used with primary move $i$ to make a rail with topmost vertex $u$.
Lines~6--8 identify the vertices $u',v'$ that are a cross move $j$ away
from vertices $u, v$, respectively, and the edge $e'=(u',v')$ between them.
Line~9 ensures that the rail $(e,e')$ is present in $G$, that is, the primary moves
are there and the cross moves are not.
Line~10 therefore adds to $D$ the rails that have $u$ as the topmost vertex, which are 
rails of the form $((v_0, v_1), (v_2, v_3))$ such that
$u \in \{v_0, v_1, v_2, v_3\}$ and $\cell(u) =\min \{\cell(v_i)\mid 0 \leq i < 4\}$.

Line~2 of Algorithm~\ref{algorithm.find} takes $O(1)$ time when $D$ is implemented as an array.
The for-loop on Lines~3--13 has $n^2$ iterations.
The for-loop on Lines~4--11 has at most $8$ iterations since $G$ is the subgraph of a knight's
graph which has degree 8.
The for-loop on Lines~5--12 has $4$ iterations.
Lines~6--8 take $O(1)$ time since function $\dest$ can be computed in $O(1)$ time.
Line~9 takes $O(1)$ time when $G$ is implemented as an adjacency list.
Line~10 takes $O(1)$ time if we append $(e,e')$ to the end of the array implementation of $D$.
Therefore, function {\sc FindRails} runs in $O(n^2)$ time.
\end{proof}

\begin{figure}[H]
  \centering
    \includegraphics[scale=0.6]{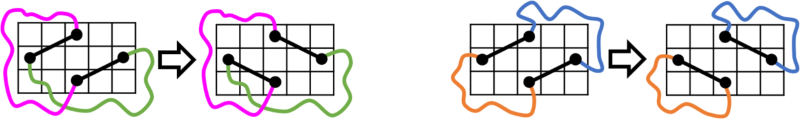}
  \caption{Switching a rail in a closed knight's tour may result in a closed knight's tour
	(left) or a pair of cycles (right).
	The curved lines indicate a sequence of knight's moves on disjoint cells.}
  \label{fig:switch}
\end{figure}

A rail may be {\em switched\/} by deleting its edges and replacing them with the complementary pair of edges. This operation preserves degree of the graph and therefore switching a rail in a closed knight's tour results in either a single closed knight's tour as shown in Figure~\ref{fig:switch} (left), or two of them as shown in Figure~\ref{fig:switch} (right).

\noindent
All of the knight's tours that we have examined to date have a large number of rails. We therefore make the following conjecture:

\begin{conjecture}
(The Rail Conjecture)
An $n \times n$ knight's tour has $\Omega(n)$ rails.
\end{conjecture}


\section{The Join Algorithm}
\label{sec:join}

The {\em rail graph\/} of a tourney is a multigraph that has a vertex for each knight and an edge between vertices $u,v$ for each rail that has one move from  knight $u$ and the from knight $v \not= u$. 
Switching the rails corresponding to the edges in a spanning forest of $G$ will almost always result in a smaller tourney\footnote{Continuing the jousting analogy, some of the knights are unhorsed and must withdraw.}, and very often a closed knight's tour. We call this the {\em join\/} operation, described more formally in Algorithm~\ref{algorithm.join}.

\begin{algorithm}[htb]
\begin{algorithmic}[1]
\Function{Join}{$G$} \Comment $G$ is a cycle cover of $K_n$ for some $n \geq 6$
\State $S \gets \mbox{\sc FindRails}(G)$ \Comment Algorithm~\ref{algorithm.find}
\State Construct the rail graph $R$ of $G$ using $S$
\State Find a spanning forest $T$ of $R$
\State Let $D$ be the set of rails corresponding to the edges of $T$
\For{$r \in D$}
  \State Switch $r$
\EndFor
\State\Return $G$
\EndFunction
\end{algorithmic}
\caption{The tourney join operation.}
\label{algorithm.join}
\end{algorithm}

\begin{figure}[htb]
  \centering
    \includegraphics[scale=0.6]{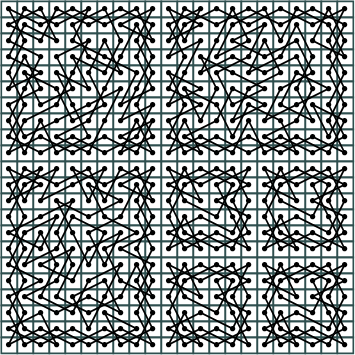}
		\hspace{0.1in}
    \includegraphics[scale=0.6]{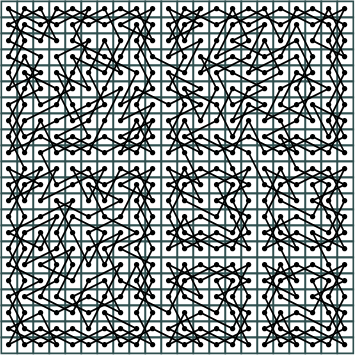}
  \caption{A $22 \times 22$ tourney of size 7 (left), and a closed knight's tour obtained from it by switching 6 vertex-disjoint rails from the spanning tree of its rail graph shown in Figure~\ref{fig:spanningtree}.}
  \label{fig:cover}
\end{figure}

\begin{figure}[H]
  \centering
    \includegraphics[scale=0.6]{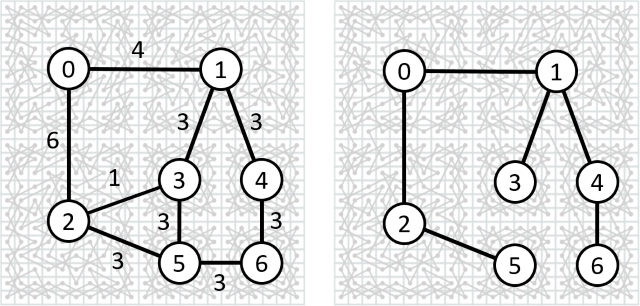}
  \caption{The rail graph of the cycle cover in Figure~\ref{fig:cover} (left). The numbers on the edges indicate the number of rails that intersect exactly two cycles. The breadth-first spanning tree of this graph is shown on the right.} 
  \label{fig:spanningtree}
\end{figure}

\begin{theorem}
\label{thm:join}
If $G \in \mathfrak{T}_n$ has size $k$, then {\sc Join}$(G)$ returns a tourney of size at most $k$ in time $O(n^2)$.
\end{theorem}

\begin{proof}
Suppose $G \in \mathfrak{T}_n$ is implemented as an adjacency list. 
Consider Algorithm~\ref{algorithm.join}.
The rail graph $R$ of $G$ has at most $n^2/4$ vertices and $O(n^2)$ edges. 
Therefore, an adjacency-list representation of of $R$ can be constructed in $O(n^2)$
 time in line~2. 
Since $R$ has $O(n^2)$ edges, a spanning tree $T$ of $R$ can be found in $O(n^2)$ time  
in line~3 using, for example, depth-first or breadth-first search.
Since $T$ has at most $n^2/4$ vertices and at most $n^2/4 - 1$ edges, $D$ can be constructed in time $O(n^2)$ in line~4 using, for example, a pre-order traversal of $T$. Since $|D| \leq n^2/4 - 1$,
the loop on lines~5--6 iterates fewer than $n^2/4$ times,
and once again the rail switch in line~6 takes $O(1)$ time.
{\sc Join}($G$) can therefore be implemented in $O(n^2)$ total time.
Clearly if $G$ is a tourney of size $k$ and the spanning tree of its rail graph $T$ has $m > 0$ edges, then {\sc Join}($G$)
will be a tourney of size $k - m$.
\end{proof}

\noindent
In practice Algorithm~\ref{algorithm.join} will generally create a knight's tour from a tourney, but it may fail to do so on occasion, particularly on small boards.


\section{Tourney Generation}
\label{sec:tourneys}

Following Tutte~\cite{tutte1954short}, the problem of finding a cycle cover of the knight's graph $K_n$ can be reduced in $O(n^2)$ time to the problem of finding a maximum cardinality matching in an undirected bipartite graph with $7n^2 - O(n)$ vertices and
$56n^2 - O(n)$ edges. Even the most efficient algorithm to date for maximum cardinality matching due to Micali and Vazirani~\cite{MicaliVazirani1980} requires $\Theta(n^3)$ time to run which, combined with the overhead involved in its implementation, makes it impractical as a method for generating tourneys.

However, tourneys are relatively easy to construct and can be converted into knight's tours using Algorithm~\ref{algorithm.join}.
For example, Figure~\ref{fig:cover} (left) shows a $22 \times 22$ tourney of size 7 constructed
using the divide-and-conquer algorithm of Parberry~\cite{ParberryKnight1997} without joining the small tours in the base of the recursion. 
Figure~\ref{fig:cover} (right) shows a knight's tour obtained from it using Algorithm~\ref{algorithm.join}.

\begin{figure}[htb]
  \centering
    \includegraphics[scale=0.6]{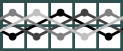}
  \caption{Portion of a horizontal braid.}
  \label{fig:braid}
\end{figure}

\begin{figure}[htb]
  \centering
    \includegraphics[scale=0.4]{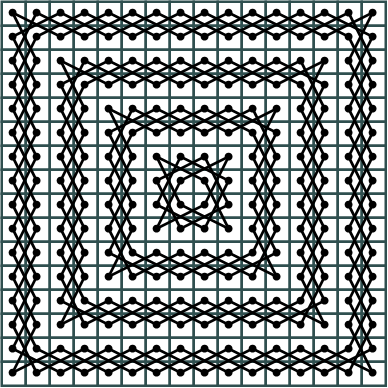}
	  \hspace*{0.1in} 
    \includegraphics[scale=0.4]{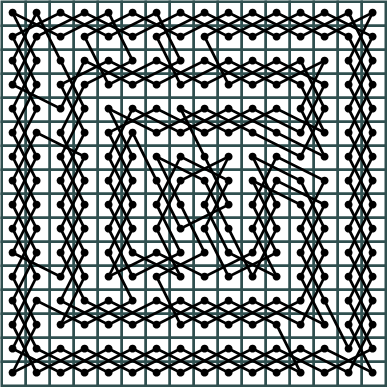}
  \caption{The $16 \times 16$ concentric braided tourney (left) and a knight's tour constructed from it  using Algorithm~\ref{algorithm.join} (right).}
  \label{fig:braided1}
\end{figure}

Some interesting tourneys may be constructed as follows.
A {\em braid\/} consists of four interwoven cycles on the knight's graph, a portion of which is shown in Figure~\ref{fig:braid}. Braid fragments often appear along the edges of knight's tours generated using Warnsdorff's algorithm (see, for example Figure~\ref{fig:warnsdorff32}).
For all even $n \geq 4$, an $n \times n$ tourney of size $4 \lfloor n/4 \rfloor$,
which we will call a {\em concentric braided tourney}, can be constructed from $\lfloor n/4 \rfloor - 1$ concentric braids around an $m \times m$ center, where $m = 4 + (n \bmod 4) \in \{4,6\}$.
For example, Figure~\ref{fig:braided1} shows $8 \times 8$ and $10 \times 10$ concentric braided tourneys.

\begin{figure}[htb]
  \centering
    \includegraphics[scale=0.4]{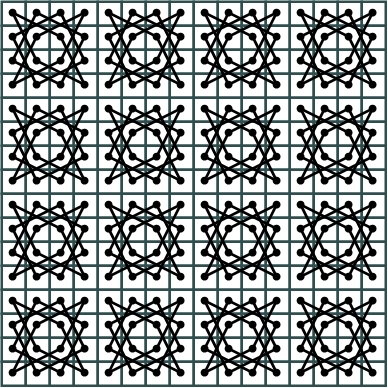}
	  \hspace*{0.1in} 
    \includegraphics[scale=0.4]{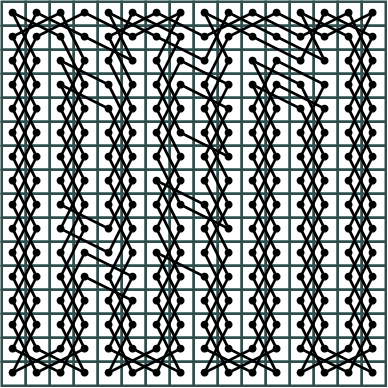}
  \caption{The $16 \times 16$ four-cover tourney (left) and a knight's tour constructed from it using Algorithm~\ref{algorithm.join} (right).}
  \label{fig:4cover1}
\end{figure}

\begin{figure}[htb]
  \centering
    \includegraphics[scale=0.44]{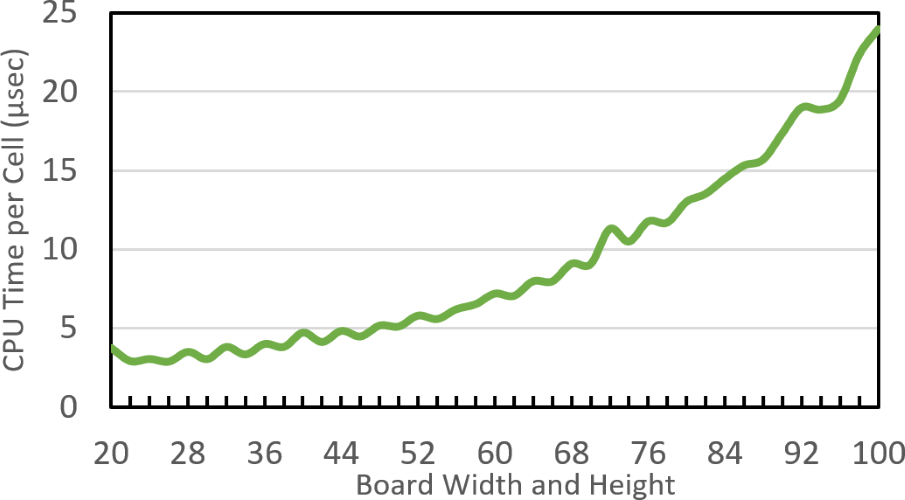}
		
  \caption{CPU time per cell averaged over 1,000 $n \times n$
		tourneys generated by Warnsdorff's algorithm for even $20 \leq n \leq 100$.}
  \label{fig:time-wy}
\end{figure}

Tourneys can also be generated by a variant of Warnsdorff's algorithm that closes off each random walk that lands in a cell that is one knight's move away from the start of that walk, and then begins a new random walk instead of starting again.
We performed experiments that measured the CPU time required to generate 1,000 tourneys on $K_n$ for even $n$ such that $20 \leq n \leq 100$. The results can be seen in Figure~\ref{fig:time-wy}. The tourney algorithm has a clear advantage, and by $n=50$ was over 200 times faster then the knight's tour algorithm (compare to Figure~\ref{fig:time-wt}).
Takefuji and Lee's neural network~\cite{Takefuji92Neural} also runs much faster than reported by Parberry~\cite{ParberryKnight1996} if it is allowed to generate tourneys instead of knight's tours.

\section{Obfuscation of Knight's Tours}
\label{sec:shatter}

To {\em shatter\/} a knight's tour, switch a randomly selected set $D$ of pairwise-disjoint rails, that is, no two distinct rails in $D$ have a vertex in common.
Shattering a closed knight's tour will in general result in a tourney.

\begin{algorithm}[htb]
\begin{algorithmic}[1]
\Function{Shatter}{$G$} \Comment $G$ is a subgraph of $K_n$ for some $n \geq 6$
\State {Construct a maximal set $D$ of disjoint rails in $G$}
\For{each rail $r \in D$}
  \State Switch $r$
\EndFor
\State\Return $G$
\EndFunction
\end{algorithmic}
\caption{Shattering a subgraph of the knight's graph.}
\label{algorithm.break}
\end{algorithm}

\begin{theorem}
\label{thm:shatter}
If $G \in \mathfrak{C}_n$, then {\sc Shatter}$(G)$ returns a tourney in time $O(n^2)$.
\end{theorem}

\begin{proof}
Suppose $G = (V, E)$ is a closed knight's tour on $K_n$ implemented as an adjacency list. 
Consider Algorithm~\ref{algorithm.break}.
Line~2 can be implemented in time $O(n^2)$ as using a straightforward linear scan of $V$
since, by Lemma~\ref{lemma:railcount}, each vertex can be a part of $O(1)$ rails.  
The loop on lines~3--4 iterates $O(n^2)$ times since $|D| = O(n^2)$,
and the rail switch in line~4 takes $O(1)$ time. 
{\sc Shatter}($G$) can therefore be implemented in $O(n^2)$ total time. Since each rail switch either preserves a cycle or splits it into two cycles, the resulting graph is a tourney.
\end{proof}

\begin{figure}[hbt]
  \centering
    \includegraphics[scale=0.44]{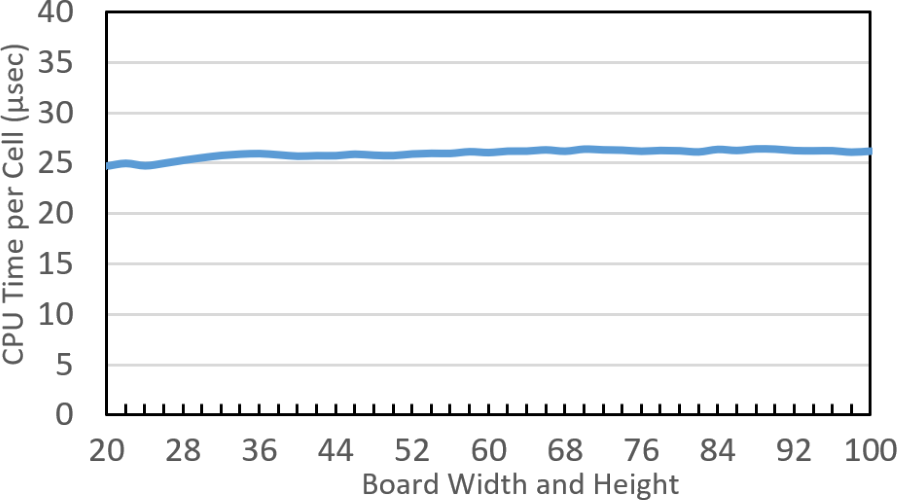}
  \caption{CPU time per cell averaged over 1,000 $n \times n$ obfuscated knight's
		tours generated by divide-and-conquer for even $20 \leq n \leq 100$.}
  \label{fig:blurtime}
\end{figure}

\begin{figure}[hbt]
  \centering
    \includegraphics[scale=0.3]{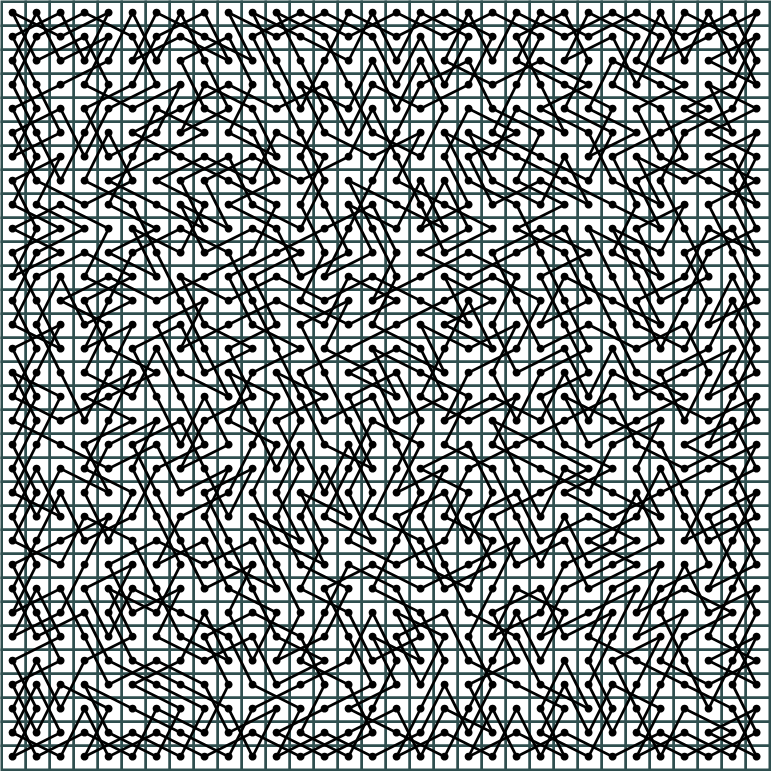}
	  \hspace*{0.02in} 
    \includegraphics[scale=0.3]{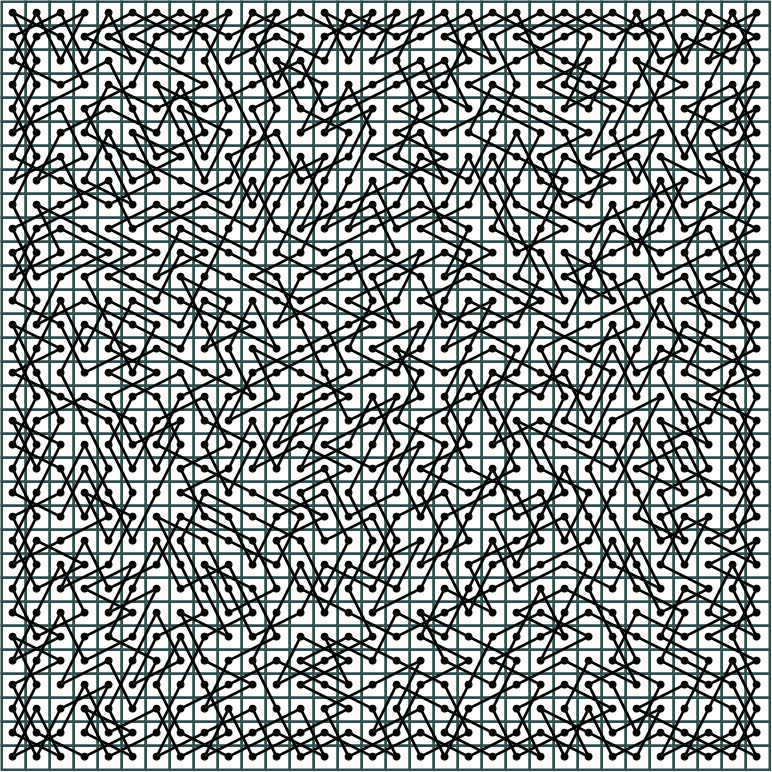}\\
		\vspace*{0.1in} 
    \includegraphics[scale=0.3]{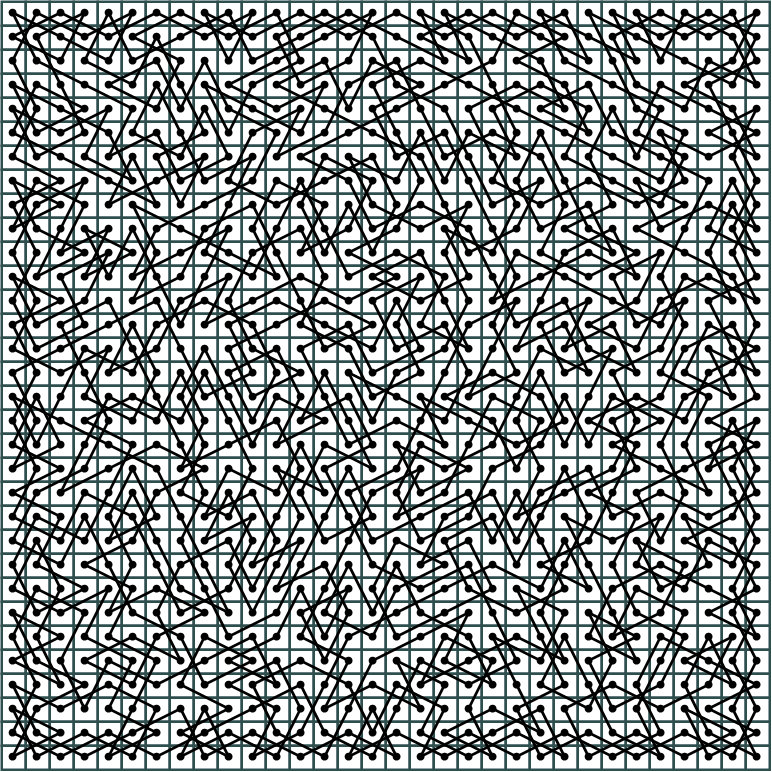}
	  \hspace*{0.02in} 
    \includegraphics[scale=0.3]{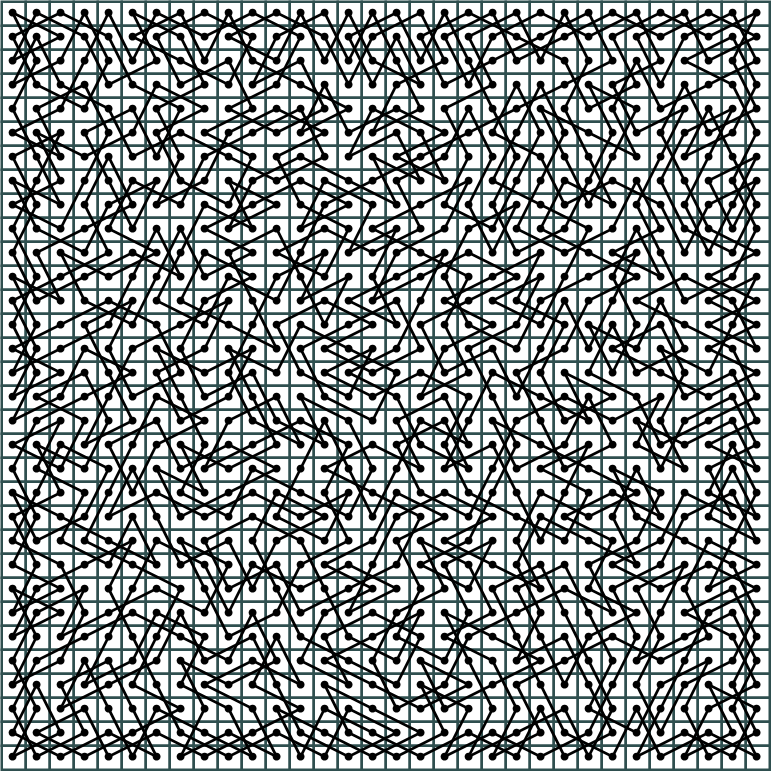}
  \caption{Obfuscated $32 \times 32$ knight's tours from (in row-major order) Warnsdorff's algorithm, divide-and-conquer, a braided tourney, and a four-cover tourney.}
  \label{fig:blurred22}
\end{figure}

\begin{table}[H]
\centering\small
\begin{tabular}{|l|c||*{8}{c|}}
\hline
\multicolumn{1}{|c|}{\bf Algorithm} & {\bf Size} &
{\bf 0} & {\bf 1} & {\bf 2} & {\bf 3} & {\bf 4} & {\bf 5} & {\bf 6} & {\bf 7}\\
\hline
\hline
Warnsdorff & $50 \times 50$ & 0.1249 & 0.1248 & 0.1250 & 0.1251 & 0.1252 & 0.1249 & 0.1245 & 0.1256\\
Takefuji-Lee & $40 \times 40$ & 0.1255 & 0.1249 & 0.1247 & 0.1252 & 0.1250 & 0.1250 & 0.1252 & 0.1244\\
Div-and-Conq & $50 \times 50$ & 0.1214 & 0.1210 & 0.1289 & 0.1286 & 0.1217 & 0.1208 & 0.1289 & 0.1287\\
Braid & $50 \times 50$ & 0.1253 & 0.1250 & 0.1251 & 0.1246 & 0.1253 & 0.1249 & 0.1253 & 0.1245
\\
Four-Cover & $48 \times 48$ & 0.1249 & 0.1248 & 0.1250 & 0.1251 & 0.1252 & 0.1249 & 0.1245 & 0.1256\\
\hline
\end{tabular}
\caption{Move distribution for 1,000 obfuscated knight's tours.}
\label{table:blur1}
\end{table}

\begin{table}[H]
\centering\small
\begin{tabular}{|l|c||*{7}{c|}}
\hline
\multicolumn{1}{|c|}{\bf Algorithm} & {\bf Size} &
{\bf 0} & {\bf 1} & {\bf 2} & {\bf 3} & {\bf 5} & {\bf 6} & {\bf 7}\\
\hline
\hline
Warnsdorff & $50 \times 50$ & 0.1444 & 0.1591 & 0.1345 & 0.1334 & 0.1339 & 0.1361 & 0.1587\\
Takefuji-Lee & $40 \times 40$ & 0.1514 & 0.1488 & 0.1359 & 0.1406 & 0.1390 & 0.1353 & 0.1490\\
Div-and-Conq & $50 \times 50$ & 0.1494 & 0.1527 & 0.1379 & 0.1356 &	0.1350 & 0.1368 & 0.1527\\
Braid & $50 \times 50$ & 0.1570 & 0.1485 & 0.1355 & 0.1374 & 0.1387 & 0.1352 & 0.1477\\
Four-Cover & $48 \times 48$ & 0.1444 & 0.1591 & 0.1345 & 0.1334 & 0.1339 & 0.1361 & 0.1587\\
\hline
\end{tabular}
\caption{Relative move distribution for 1,000 obfuscated knight's tours.} 
\label{table:blur2}
\end{table}

\pagebreak

To {\em obfuscate\/} a knight's tour, shatter it with Algorithm~\ref{algorithm.break} a small constant number of times, then join it with Algorithm~\ref{algorithm.join}. This
(by Theorems~\ref{thm:join} and~\ref{thm:shatter}) takes $O(n^2)$ time, that is, a constant amount of time per cell.
Sixteen iterations of shatter were sufficient for the examples used in this section. 
The CPU time per cell for generating and obfuscating an $n \times n$ divide-and-conquer knights tour for even $20 \leq n \leq 100$, shown in Figure~\ref{fig:blurtime}, is consistent with this claim. 
Figure~\ref{fig:blurred22} shows four obfuscated knight's tours that look very similar in spite of being generated by four very different algorithms.
Table~\ref{table:blur1} shows the move distribution for obfuscated knight's tours generated by five different algorithms. The standard deviation was less than 0.001 in each case. 
Table~\ref{table:blur2} shows the corresponding relative move distribution. The standard deviation was again less than 0.001 in each case. More information can be found in the Supplementary Material (see Section~\ref{sec:supplement}).


\section{Conclusion}
\label{sec:conclusion}

We have introduced the concept of a {\em tourney}, which is a vertex-disjoint cycle cover of the knight's graph, and described several methods of generating them. Using the concept of a {\em rail\/} consisting of a pair of vertex-disjoint moves on four adjacent vertices of the knight's graph, we have shown how to {\em join\/} tourneys into closed knight's tours using a spanning tree of a multigraph called the {\em rail graph}. 
With an algorithm for {\em shattering\/} knight's tours into tourneys, this gives a method for {\em obfuscating\/} closed knight's tours to obscure visual artifacts caused by their method of generation. We have provided visual and statistical evidence of the efficacy of our obfuscation algorithm.  
Open problems include a proof (or counterexample to) the Rail Conjecture (see Section~\ref{sec:rails}).


\section{Supplementary Material}
\label{sec:supplement}

Supplementary material including
the run-time and move distribution data exhibited above  
and additional images that are too large for this paper can be browsed online at:
\begin{center}
  \url{http://ianparberry.com/research/tourneys/}.
\end{center}

\noindent
Open source, cross platform C++ code for the tourney generator that was used to generate the images and data for this paper can be cloned or downloaded from:
\begin{center}
  \url{https://github.com/Ian-Parberry/Tourney}.
\end{center}
This generator outputs tourneys in Scalable Vector Graphics (SVG) format which can be viewed in a web browser, and also in text format suitable for input to any program that the user may wish to write. It will also generate run-time and move distribution data in a text file that can be imported into a spreadsheet.
For more details, see the Doxygen-generated code documentation at:
\begin{center}
  \url{http://ianparberry.com/research/tourneys/doxygen/}.
\end{center}


\end{document}